\documentclass[letterpaper,twocolumn,superscriptaddress,floatfix]{revtex4}
\usepackage[latin1]{inputenc}
\usepackage{bm}
\usepackage{multirow,amssymb,amsbsy,amsmath}
\usepackage{graphicx}
\usepackage{verbatim}
\makeatletter
\usepackage{pifont}
\makeatother

\begin{document}

\title{Storage of multiple single-photon pulses emitted from a quantum dot in a solid-state quantum memory}

\author{Jian-Shun Tang}
\thanks{These authors contributed equally to this work.}
\author{Zong-Quan Zhou}
\thanks{These authors contributed equally to this work.}
\author{Yi-Tao Wang}
\author{Yu-Long Li}
\author{Xiao Liu}
\author{Yi-Lin Hua}
\author{Yang Zou}
\author{Shuang Wang}
\author{De-Yong He}
\author{Geng Chen}
\author{Yong-Nan Sun}
\affiliation{Key Laboratory of Quantum Information, University of Science and Technology of China, CAS, Hefei, Anhui 230026, China.}
\affiliation{Synergetic Innovation Center of Quantum Information $\&$ Quantum Physics, University of Science and Technology of China, Hefei, Anhui 230026, China.}

\author{Ying Yu}
\author{Mi-Feng Li}
\author{Guo-Wei Zha}
\author{Hai-Qiao Ni}
\author{Zhi-Chuan Niu}
\email{zcniu@semi.ac.cn}
\affiliation{Synergetic Innovation Center of Quantum Information $\&$ Quantum Physics, University of Science and Technology of China, Hefei, Anhui 230026, China.}
\affiliation{The state key laboratory for superlattices and microstructures, Institute of semiconductors, CAS, PO Box 912, Beijing 100083.}

\author{Chuan-Feng Li}
\email{cfli@ustc.edu.cn}
\author{Guang-Can Guo}
\affiliation{Key Laboratory of Quantum Information, University of Science and Technology of China, CAS, Hefei, Anhui 230026, China.}
\affiliation{Synergetic Innovation Center of Quantum Information $\&$ Quantum Physics, University of Science and Technology of China, Hefei, Anhui 230026, China.}

\date{\today}

\begin{abstract}
Quantum repeaters are critical components for distributing entanglement over long distances in presence of unavoidable optical losses during transmission. Stimulated by Duan-Lukin-Cirac-Zoller protocol, many improved quantum-repeater protocols based on quantum memories have been proposed, which commonly focus on the entanglement-distribution rate. Among these protocols, the elimination of multi-photons (multi-photon-pairs) and the use of multimode quantum memory are demonstrated to have the ability to greatly improve the entanglement-distribution rate. Here, we demonstrate the storage of deterministic single photons emitted from a quantum dot in a polarization-maintaining solid-state quantum memory; in addition, multi-temporal-mode memory with $1$, $20$ and $100$ narrow single-photon pulses is also demonstrated. Multi-photons are eliminated, and only one photon at most is contained in each pulse. Moreover, the solid-state properties of both sub-systems make this configuration more stable and easier to be scalable. Our work will be helpful in the construction of efficient quantum repeaters based on all-solid-state devices.
\end{abstract}

\maketitle

Long-distance entanglement distribution has become increasingly important, which is essential in the improvement of many quantum technologies, such as quantum key distribution \cite{gisin2002} and quantum internet \cite{kimble2008}. It is also helpful in the examination of the foundation problems in quantum mechanics, for example, the Bell-inequality test \cite{bell1964}. However, this task is not easy to perform, because of the photon loss during the fiber transmission. One proposal to overcome this issue is to use quantum repeaters \cite{briegel1998}. In this architecture, the entire distance is divided into several shorter elementary links, and for each link, entanglement between quantum memories can be established independently. Next, the elementary links are joined using entanglement swapping to create an entangled pair over the entire distance.

The first concrete quantum repeater proposal is the well-known Duan-Lukin-Cirac-Zoller (DLCZ) protocol \cite{duan2001}, in which the atomic ensembles are used as the quantum memories and the photon sources. However, the entanglement distribution rate is still quite low in this scheme. To solve this problem, many improved architectures of quantum repeaters have been proposed \cite{simon2007,sangouard2007,sangouard2008,sinclair2014,krovi2015}. Of these proposals, the main approaches include the single photon source \cite{sangouard2007} (or the single entangled photon-pair source) and the multimode quantum memory \cite{simon2007,sinclair2014}, etc. Using a single photon source, the multiphoton errors can be eliminated, which makes the photon emission rate can be greatly enhanced, thus improving the entanglement distribution rate. Meanwhile, the multimode-memory protocol has been estimated to be the most efficient protocol among all the quantum repeater architectures \cite{sangouard2011rmp}.

On the one hand, quantum dots (QDs) are attracting increasing interest in the quantum memory research community, and notable recent work includes the slowing-down experiment of QD emissions in the hot atomic ensemble \cite{akopian2011}, and the absorption of QD emissions in a single ion \cite{meyer2014}. This increasing interest is primarily because of the potential of QDs to be a good source of deterministic single photons \cite{tang2012,he2013} and also a good source of deterministic entangled photon pairs \cite{muller2014}. On the other hand, multimode-quantum-memory protocols are being intensively studied. Many degrees of freedom of the photons, such as time bins \cite{usmani2010,clausen2011,saglamyurek2011,saglamyurek2015,tiranov2015}, frequencies \cite{sinclair2014}, polarizations \cite{tiranov2015,clausen2012,gundogan2012,zhou2012,bussieres2014}, spatial directions \cite{lan2009} and orbital-angular-momentum (OAM) modes \cite{zhouLGMM}, can be used to multiplex the memory and the communication channel. Time bins are one type of the most-used modes. Rare-earth (RE)-ion-doped solids have the advantage of broad inhomogeneous absorption, which provides a large memory bandwidth for these solid-state quantum memories \cite{saglamyurek2011}, and enables them to be used for temporal-multimode operations.

In this work we experimentally demonstrate two points. The first point is the storage of deterministic single photons (with no multi-photons in principle) emitted from a semiconductor self-assembled QD in a solid-state polarization-maintaining quantum memory \cite{zhou2012}, which is based on Nd$^{3+}$:YVO$_{4}$ crystals. The QD and the RE-ion-doped crystals are separated by $5$ m on two separate optical tables and are connected via a $10$-m fiber (see Figure 1). The second point is the realization of the temporal multiplexed quantum memory with QD-based narrow single-photon pulses. $1$, $20$ and $100$ temporal modes are respectively stored in the quantum memory, with at most one photon present in each mode. Both of these points will be helpful in the development of quantum repeaters. Moreover, both sub-systems in our experiment are solid-state, which will make this configuration more stable and convenient.

\begin{figure*}[tb]
\centering
\includegraphics[width=0.9\textwidth]{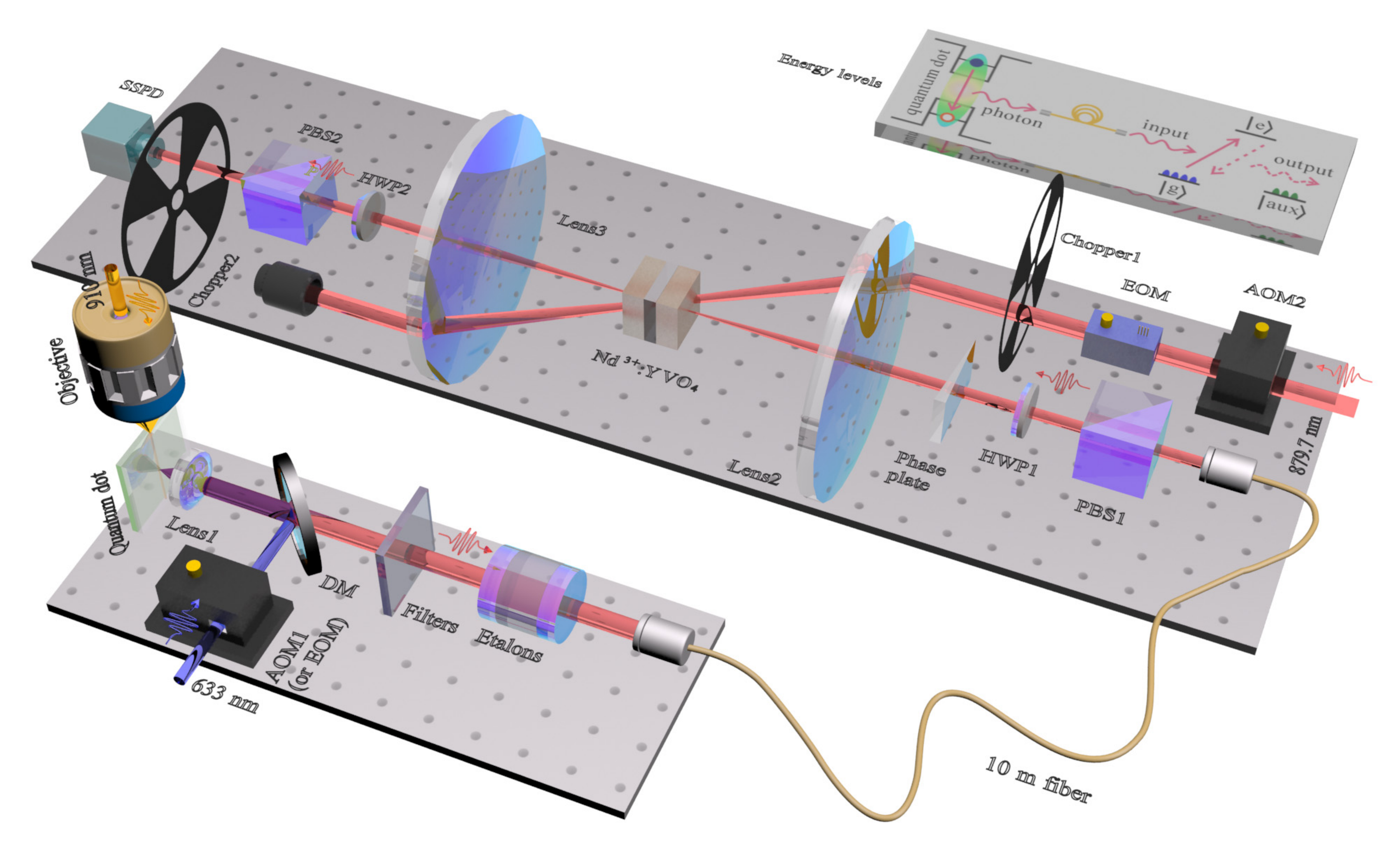}
\renewcommand{\figurename}{Figure}
\caption{\label{Fig1} \textbf{Experimental setup.} This experiment is performed on two separated optical tables connected by a $10$-m-long optical fiber. The QD sample and the Nd$^{3+}$:YVO$_{4}$ crystals are spaced $5$ m apart. A single QD embedded in a planar DBR microcavity is excited by a $633$-nm laser, and a $910$-nm laser is used to shift the wavelength of the QD emission using the local-heating effect. The QD emission is precisely shifted to the $^{4}$I$_{9/2}\rightarrow^{4}$F$_{3/2}$ transition of Nd$^{3+}$ with the calibrated etalon. The $879.7$-nm laser, which is modulated by AOM2 and EOM both in intensity and in frequency, is used to pump the Nd$^{3+}$:YVO$_{4}$ crystals to create a frequency comb according to the AFC protocol. The single photons are then stored in the crystals and retrieved after a time $T_{\rm storage}$. An SSPD with a low dark count is used to detect the single photons. PBS1, HWP1, PBS2, HWP2 and the phase plate are used to prepare and measure the polarization qubit when the qubit-memory experiment is performed. The arrows indicate the directions of the light beams. The AOMs (or EOM used for modulating excitation light) and the choppers are synchronized to an electrical-pulse generator to create the time sequence for this experiment. The inset shows the energy levels of the QD and the Nd$^{3+}$-ions. The QD, which can contain an exciton, biexciton or a trion, emits a single photon, which is then sent to the quantum memory via a fiber, and is subsequently absorbed by the ion ensemble with a frequency comb. After the storage time, the photon is re-emitted by the ions. $|g\rangle$, $|e\rangle$ and $|aux\rangle$ denote the ground, excited and auxiliary levels of Nd$^{3+}$ ions, respectively.}
\end{figure*}

\textbf{Results}

\textbf{Wavelength matching between QD and the quantum memory.} The absorption line of the Nd$^{3+}$:YVO$_{4}$ crystal is located at $879.7$ nm, with a narrow memory bandwidth of hundreds of MHz, whereas the emission lines of InAs/GaAs QDs are random, depending on the dot size and the chemical composition in a wide range of approximately $900\sim1200$ nm. $879.7$ nm is an embarrassing wavelength because it is far away from the emission range of QDs. Moreover, the wetting layer emission of the InAs/GaAs QD sample is located at approximately $860$ nm, with a full width of half maximum (FWHM) of tens of nm, and is much more intense than the signal photons. This emission represents a strong noise source at $879.7$ nm. To solve this problem, we employ three steps, including QD sample fabrication with a special design (blue-shift the QD emissions to around the memory band, see Methods and Supplementary Note 1 for details, and the sample structure is shown in Supplementary Figure 1), QD selection (select an emission nearest to the memory band, see Methods) and local heating of the selected QD with a strong laser (finely tune the QD emission to the memory band, see Methods), to derive a QD emission that matches the memory band of the quantum memory.

\begin{figure}[tb]
\centering
\includegraphics[width=0.45\textwidth]{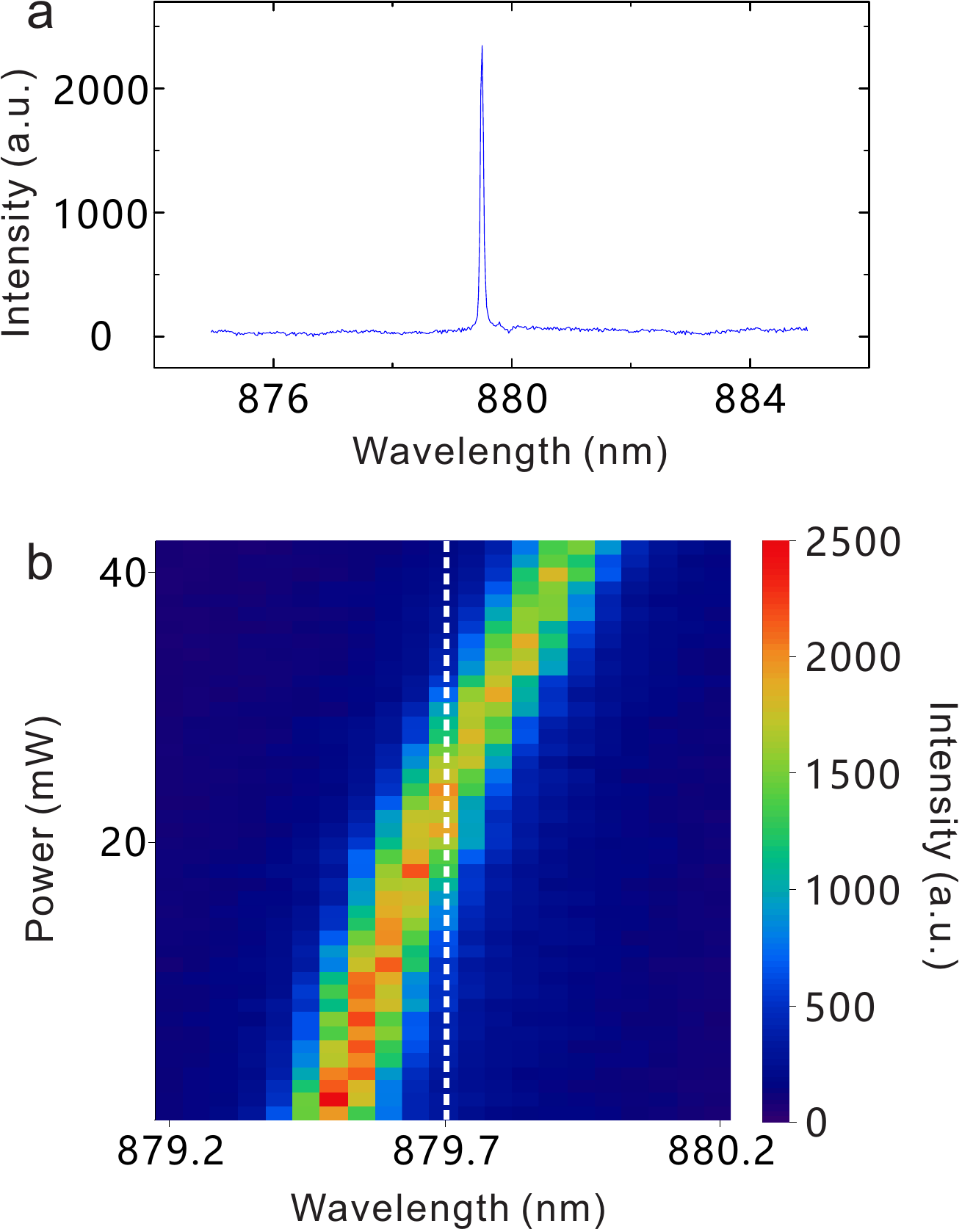}
\renewcommand{\figurename}{Figure}
\caption{\label{Fig2} \textbf{Photoluminescence spectra.} (a) Photoluminescence spectrum without local heating. A single peak appears at the position of $879.5$ nm, which may be caused by the unbalanced carrier capture and the filtering effect of the DBR microcavity (see Supplementary Note 2 for the explanations). The wavelength of this peak is a little shorter than that of the memory band ($879.7$ nm), which provides the chance to tune it to match the memory band using local heating effect. (b) Power-dependent spectra. The x-axis is the wavelength of QD emission, and the y-axis is the power of the $910$-nm laser, which is used for local heating. The color represents the photoluminescence intensity of the QD emission. The peak is shifted across $879.7$ nm with increasing laser power. When the power reaches $24$ mW, the emission line matches the memory band.}
\end{figure}

\begin{figure}[tb]
\centering
\includegraphics[width=0.4\textwidth]{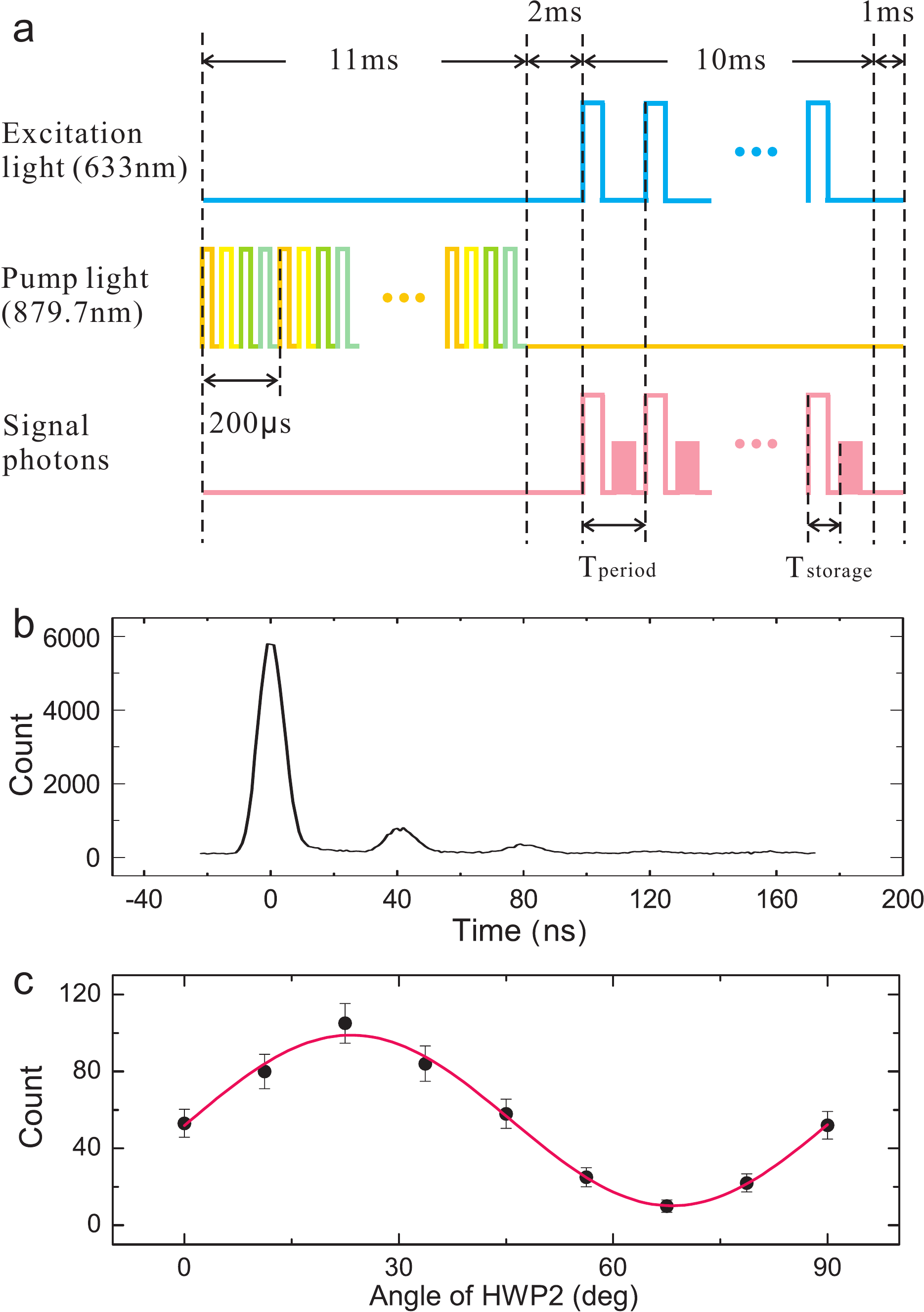}
\renewcommand{\figurename}{Figure}
\caption{\label{Fig3} \textbf{Time sequence and experimental results.} (a) Time sequence diagram. The entire procedure includes a $11.5$-ms preparation time, a $2.5$-ms wait time, a $10$-ms storage and retrieval time and another $1$-ms wait time. During the preparation time, the pump pulses with different frequencies (represented by different colors) are used to pump the Nd$^{3+}$:YVO$_{4}$ crystals to prepare the AFC (see Supplementary Note 4 for methodological details). In the storage and retrieval procedure, the pump laser is blocked, and then the excitation light associated with the single photons is modulated to a series of pulses with a period of $T_{\rm period}$. After a storage time $T_{\rm storage}$, the signal photons are retrieved. (b) Time spectrum of the stored single photons. The storage time is $40$ ns, and the pulsewidth of the excitation light is $10$ ns. The first peak represents the light that is not absorbed. The second peak represents the retrieved single photons, and the third peak is the second-order retrieved photons. (c) Polarization-qubit quantum memory. Single photons are encoded with a qubit $|H\rangle+|V\rangle$ and then sent to the quantum memory. The storage time is $40$ ns, and a $2$ ns coincidence window is chosen for the retrieved single photons. The retrieved qubit is projected to a series of bases, which are represented by the angles of HWP2. The count shows a sinusoidal oscillation with a fidelity of $0.913\pm0.026$. The error bars in these data are due to the counting statistics.}
\end{figure}

\begin{figure}[tb]
\centering
\includegraphics[width=0.4\textwidth]{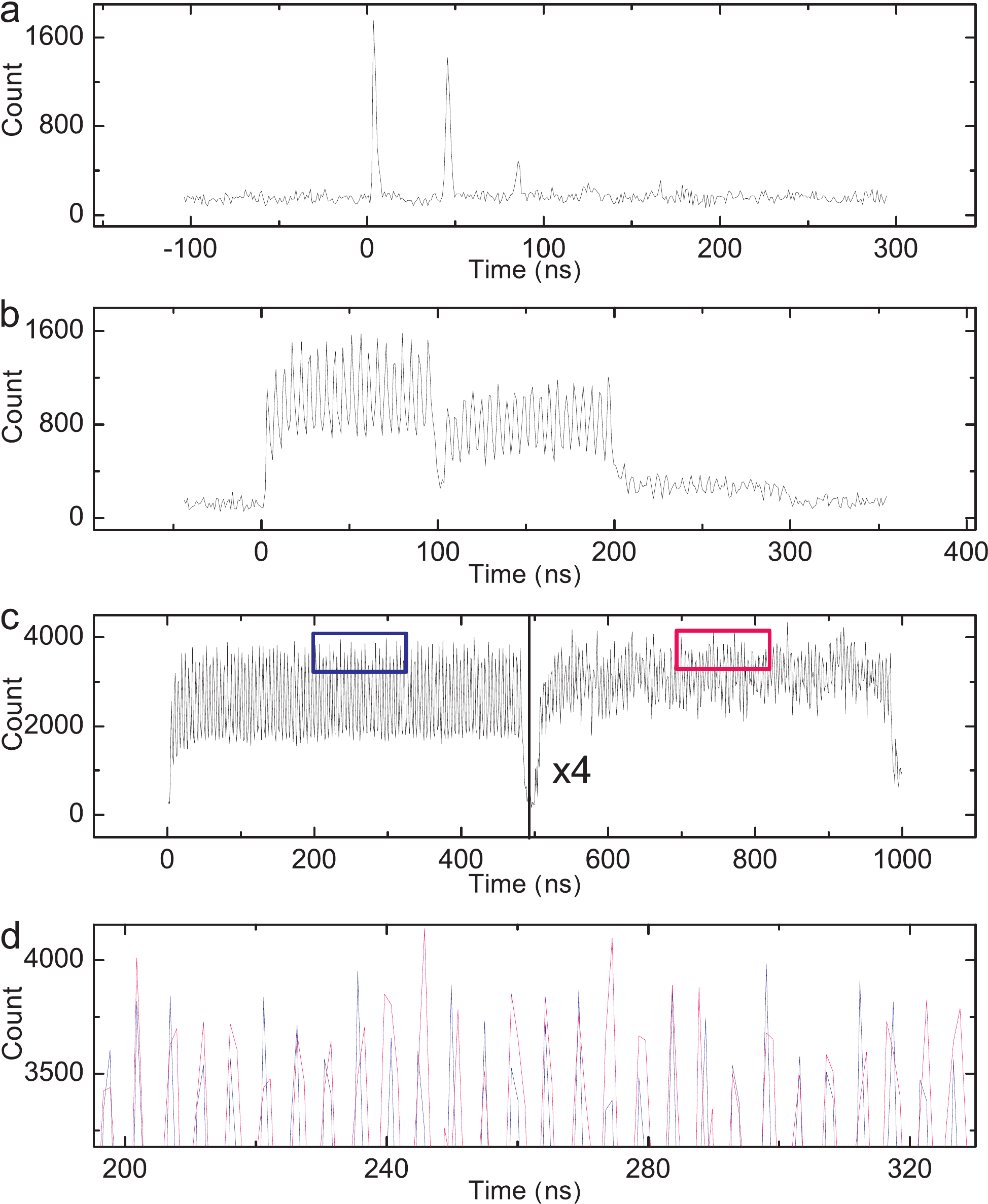}
\renewcommand{\figurename}{Figure}
\caption{\label{Fig4} \textbf{The quantum storage of multiple single-photon pulses.} We use an EOM to modulate the excitation light here, and its pulsewidth $T_{\rm expw}$ is reduced to $0.8$ ns, which ensures there is at most one photon in each pulse. (a) $1$, (b) $20$ and (c) $100$ temporal modes of the single photons are used for the quantum memory. The first group of peaks represents the transmitted photons, second group of peaks represents the stored single photons, and the third group of peaks (not in (c)) represents the second-order retrieved photons. The storage times are $40$ ns, $100$ ns and $500$ ns, respectively. In (c) the counts of the second group of peaks are multiplied by a factor $4$. (d) The enlargement of the rectangle-regions in (c). The blue peaks correspond to the transmitted-photon signals in the blue rectangle, and the pink peaks correspond to the stored-photon signals in the pink rectangle, but with the time coordinate subtracted by $500$ ns (the storage time). By comparing these two groups of peaks, we find that each of the peaks in the transmitted and stored signals correspond to each other well. This result shows that the temporal modes of the single photons are well maintained during the memory process.}
\end{figure}

Figure 2 shows the experimental results. Figure 2(a) is the photoluminescence spectrum of a QD which is sifted out from hundreds of QDs on this specially designed QD sample, namely, the result of the first and second steps. The reason for selecting this QD is that its emission line is at the wavelength of $879.5$ nm, which is slightly shorter than the wavelength of the memory band. However, no other lines are detected in the region. This result may be caused by the unbalanced carrier capture (this process may cause the generation of trion which can be examined by the vanish of fine structure splitting \cite{bayer1999,bayer2002}) and the filtering effect of the planar distributed Bragg reflector (DBR) microcavity (see Supplementary Note 2 and Supplementary Figure 2 for details). Figure 2(b) shows the local-heating spectrum for this QD (see Supplementary Note 3 for local heating, and the sketch of the local heating and some additional data are shown in Supplementary Figure 3). When the power of the $910$-nm heating laser reaches $24$ mW, the emission line can be finely red-shifted to $879.7$ nm \cite{gong2010}, i.e., the memory band. The autocorrelation of this emission line is measured to be approximately $g^{(2)}(0)=0.14$, which demonstrates that the photons were emitted from a good-quality single photon source.

\textbf{Storage for single photons.} In our experiment, the quantum memory is based on the atomic frequency comb (AFC) protocol, according to which the absorption band of the Nd$^{3+}$-ion ensemble in the crystals (see Methods for the details of the quantum-memory sample) must be tailored into a frequency comb. The period of the frequency comb (denoted as $\Delta$) determines the storage time of the single photons ($T_{\rm storage}$), and the range of the frequency comb determines the memory bandwidth. The details about the AFC quantum memory technique are presented in the Methods and in Supplementary Note 4, and a sketch is shown in Supplementary Figure 4.

Figure 3(b) shows the experimental result of the storage of the single photons. To obtain this time spectrum, the storage and retrieval period ($T_{\rm period}$) and the pulsewidth of the excitation pulse ($T_{\rm expw}$, which is related to the signal pulsewidth) are set to be $400$ ns and $10$ ns, respectively. The storage time is $T_{\rm storage}=40$ ns, and the integration time is $2.8$ hours. The peak at $0$ ns corresponds to the single photons that are not absorbed as well as some background wetting-layer light, the peak at $40$ ns corresponds to the single photons that are stored and retrieved, and the peak at $80$ ns is the second-order retrieved single photons. The second peak is relatively small compared with the first peak. This result is related to another merit of the quantum memory, which can be used as a good narrow-bandwidth filter \cite{lin2011}. The single photons are filtered by the filter and etalon (Figure 1) before they are sent to the quantum memory. However, the background wetting-layer light can pass through the etalon from its other transmission peaks (although every copy of the leaked wetting-layer light is little, there are many copies of them). Next, we use the quantum memory itself as a filter to make the single photons clear. Therefore, the retrieved signal contains only single photons, and the memory efficiency is not as low as that shown in this figure; this point will be discussed later. The signal-to-noise ratio (SNR) of the stored photons is estimated to be $9:1$. Approximately $25\%$ of the noise stems from the dark counts of the superconducting single-photon detector (SSPD), $25\%$ comes from the pump laser, and $50\%$ stems from the ambient light.

Our sandwich-like quantum memory has been demonstrated to be polarization-maintaining. Process tomography has shown that its fidelity can reach $0.999\pm0.002$ using the attenuated laser \cite{zhou2012}. Here, we take a typical case as an example to show the preservation of quantum coherence during the storage and retrieval process, which is important in quantum applications (including quantum repeater). Figure 3(c) shows the experimental result. The error bars in these data are due to the counting statistics. The deterministic single photons are encoded with a quantum state of $|H+V\rangle$ by PBS1 (polarizing beamsplitter) and HWP1 (half-wave plate, see Figure 1; the phase plate is used to compensate for the phase difference induced by the crystals) and are then sent to the quantum memory (see Methods for sample structure and ``$H$", ``$V$" definitions). After a storage time of $T_{\rm storage}=40$ ns, the single photons are retrieved. The polarization characteristics of these retrieved photons are examined via HWP2 and PBS2. A $2$-ns window is selected at the position of $40$ ns, and the integration time is $1000$ s. As the angle of HWP2 is rotated from $0^{\circ}$ to $90^{\circ}$, we observe a sinusoidal oscillation of the single-photon counts. The maximum value appears at the position of $22.5^{\circ}$ ($|H+V\rangle$ basis) and the minimum value appears at $67.5^{\circ}$ ($|H-V\rangle$ basis) of HWP2. The fidelity of the retrieved polarization state is estimated to be $0.913\pm0.026$ (this value is also directly associated with the fidelity of the storage process). These results show that the $|H+V\rangle$ state can be well preserved.

\textbf{Storage of multiple single-photon pulses.} To further reduce the photon numbers in a single pulse while increasing the mode numbers of the photons that can be stored, we use an electro-optic modulator (EOM) to replace AOM1 (acousto-optic modulator; see Figure 1). The pulsewidth of the excitation laser $T_{\rm expw}=0.8$ ns is reduced to be less than the QD's lifetime, which ensures there is only one photon in a single pulse. This point is also demonstrated by the Hanbury Brown-Twiss experiment (see details in Supplementary Note 5 and Supplementary Figure 5).

Figure 4(a) shows the result of the storage for $1$ single-photon pulse with $T_{\rm period}=400$ ns and $T_{\rm storage}=40$ ns. The integration time is $11.7$ hours. The peak at $0$ ns corresponds to the single photons that are not absorbed without wetting-layer light, the peak at $40$ ns corresponds to the single photons that are stored, and the peak at $80$ ns is the second-order retrieved single photons. An additional set of filter and etalon (with a different free spectral range (FSR)) are inserted in the beampath here to filter the single photons more clearly. In this situation, the second peak is almost as high as the first peak. In Figure 4(b), $20$ single-photon pulse are stored in the quantum memory with $T_{\rm period}=400$ ns and $T_{\rm storage}=100$ ns. The integration time is $7.5$ hours, and the separation between the neighboring modes is $4.8$ ns. $20$ peaks are clearly seen in the range of $100\sim200$ ns, which are the stored single-photon temporal modes. The peaks in the ranges of $0\sim100$ ns and $200\sim300$ ns are the transmitted light and the second-order retrieved light, respectively.

We also examine the situation of $100$ modes with $T_{\rm period}=1000$ ns and $T_{\rm storage}=500$ ns, as shown in Figure 4(c). The integration time and the separation between the neighboring modes are $46.1$ hours and $4.8$ ns, respectively. The peaks in the ranges of $0\sim500$ ns and $500\sim1000$ ns are the transmitted light and the retrieved single photons, respectively. In fact, the efficiency of quantum memory is related to the storage time. In the present situation, the efficiency is approximately $7\%$, whereas this value is estimated to be $20\%$ and $13\%$ in the situations of $T_{\rm storage}=40$ ns and $T_{\rm storage}=100$ ns, respectively. In spite of the decrease of the memory efficiency, we can still clearly observe $100$ small peaks from the retrieved photons. Figure 4(d) shows the details of the peaks in the blue rectangle of Figure 4(c) and those in the pink rectangle with the time-coordinate subtracted by $T_{\rm storage}=500$ ns. Each of these peaks corresponds well to each other one by one. This phenomenon shows the reliability of our experimental results.

\textbf{Discussion}

Our work is the first demonstration of the storage of the QD-based deterministic-single-photon pulse trains. Both the sub-quantum systems in this configuration, namely, the QD and the RE-ion-doped crystals, are solid-state materials. Moreover, we demonstrate that the polarization states of the single photons can be well preserved. Therefore, both the polarization states and the time bins can be used to encode the qubits. Notably, although the single-photon case is described here, the QD also has the potential to be a high-quality deterministic entangled-photon-pair source or a solid spin qubit that is entangled with a single photon \cite{greve2012}. In these situations, our configuration could be conveniently utilized. All of these characteristics make our configuration more suitable for the construction of the efficient quantum repeaters.

One possible application of our configuration is the quantum repeater protocol recently proposed by Sinclair \emph{et al.} \cite{sinclair2014}, which is based on spectral multiplexing, multimode AFC delay quantum memory, entangled photon-pair sources, Bell-state measurement and feed-forward control. This protocol can, on the one hand, decrease the requirement of the on-demand quantum memory, and on the other hand, greatly improve the entanglement distribution rate because the connections of the elementary links are processed simultaneously. The hierarchical connections are avoided in this protocol, and thus, long-distance classical communications are also avoided. In our experimental configuration, temporally multiplexed quantum memory has been demonstrated (current work), and spatially multiplexed memory (OAM modes) has also been demonstrated to be realizable \cite{zhouLGMM}. Moreover, the QD has been revealed to possess the ability to be a narrow-linewidth indistinguishable deterministic entangled photon-pair source \cite{muller2014}. Therefore, our work makes us one step closer to this high-efficiency quantum repeater architecture. Moreover, the determinacy property of the QD emissions also plays a crucial role for the improvement of the entanglement distribution rate in this protocol, which has been calculated in detail in a further work \cite{guha2014} by Guha \emph{et al.}

Another example of the application of our configuration is the quantum repeater protocol based on single photon sources \cite{sangouard2007}, which improves upon the DLCZ protocol by replacing the photon-pair sources (an equivalent protocol as the DLCZ one) with the single photon sources. In the former case, to avoid the errors caused by the multi-photon-pair emissions, the emission rate of the photon pairs must be greatly limited; whereas in the latter case, this problem is eliminated. Therefore, the entanglement distribution rate can be improved. In our work, the single-photon storage can be utilized in this protocol (when spin-wave AFC memory is integrated), and furthermore, by combining the multi-temporal-mode quantum memory \cite{simon2007}, the quantum repeater can be faster and more robust.

Our work can probably be used in other quantum technologies, for example, the quantum networks \cite{kimble2008,duan2010,northup2014}, etc. Although the quantum information flowing between QD and the quantum memory is not demonstrated here, the quantum channel between them has been demonstrated, and the polarization information can be preserved. The generation of photon-spin entanglement has been demonstrated to be realizable in the QD \cite{greve2012}. Therefore, the exchange of quantum information between the QD and the quantum memory can be in principle achieved.

To conclude, we achieve the storage of deterministic single photons emitted from a QD in a sandwich-like Nd$^{3+}$:YVO$_{4}$ quantum memory, which can preserve the polarization states of the input photons. We have also demonstrated the temporal multimode operation of the quantum memory with $1$, $20$ and $100$ narrow single-photon pulses. Only one photon exists in a single pulse at most. Our work paves the way toward the construction of high-speed quantum repeaters based on all-solid-state devices and can also be used in other quantum technologies.

\textbf{Methods}

\textbf{QD fluorescence collection.} A $633$-nm laser is used to excite the QD, as shown in Figure 1. After being reflected by a dichroic mirror (DM), the laser is focused onto the sample by an aspheric lens (Lens1). The same lens, with a high numerical aperture (N.A.) of $0.68$ and a working distance of $1.76$ mm (placed inside the cryostat), is used to collect the fluorescence emitted from the QD. Moreover, the QDs are grown in a planar microcavity comprising a 10-pair Al$_{0.9}$Ga$_{0.1}$As/GaAs DBR on top and a 32-pair DBR on bottom. This microcavity can greatly enhance the collection efficiency of the fluorescence \cite{he2013,flagg2009}. The fluorescence is then filtered by the DM and a $20$-nm bandwidth filter ($99\%$ transmission at $879.7$ nm and $10^{-7}$ at $910$ nm). When the QD spectra are detected, the fluorescence is sent to a spectrometer (not shown) equipped with a $900$ grooves$\cdot$mm$^{-1}$ grating and a Peltier-cooled InGaAs detector array; when the autocorrelation function is measured, the fluorescence is sent to a picosecond time analyzer (not shown). Otherwise, the fluorescence is sent to the subsequent optical elements in Figure 1.

\textbf{Steps for wavelength matching.}
During the QD sample growth, an AlGaAs layer is grown under the InAs QD because AlAs has a broader energy gap than GaAs, which can blue shift the QD emissions, as well as the wetting layer emission. However, problematically, the InAs QD tends to disappear and the quantum efficiency is also substantially reduced when it meets the aluminium-containing layer. Here, we design a special structure of the QD sample, which contains a $20$-nm-thick Al$_{0.2}$Ga$_{0.8}$As layer and a $10$-nm-thick GaAs layer overlying it (see Supplementary Note 1 and Supplementary Figure 1). The InAs QDs are then grown on the GaAs layer using the molecular beam epitaxy (MBE) technique. With this special design, the emission lines of QDs are blue shifted to approximately $879.7$ nm, and the emissions have a relatively high quantum efficiency and a high SNR. The QD emissions roughly coincide with the $^{4}$I$_{9/2}\rightarrow^{4}$F$_{3/2}$ transition of Nd$^{3+}$ in the YVO$_{4}$ crystal \cite{zhou2012}.

QDs have nonuniform emission wavelengths and emission strength. To find a QD with emission at a wavelength that is slightly shorter than $879.7$ nm (the wavelength can only be red shifted for several $0.1$ nm in the following fine-tuning process) and with a high brightness, we place the QD sample on a 3-axis nano-positioner with the travel ranges of several millimeters, and the entire setup is then placed in a low-vibration liquid-helium-free cryostat, where the sample is cooled to $8$ K (Montana Instruments). Next, we scan the nano-positioner in the $xy$-plane to detect the QD spectra. After scanning, we compare hundreds of QDs and then identify a single emission line located at the wavelength of $879.5$ nm.

To tune this emission line to the Nd$^{3+}$ ion transition exactly, we use a $910$-nm laser to perform local heating in the vicinity of the selected QD (see Supplementary Note 3 and Supplementary Figure 3(a)). This laser is focused to the cleaved edge of the DBR microcavity using a 50X long working distance objective (see Figure 1). The microcavity acts as a waveguide for the laser and directs it to the vicinity of the selected QD \cite{flagg2009}. The orthogonal geometry and the filter completely eliminate the residual local-heating laser. Compared with global heating, local heating can induce less noise and less inconvenience due to heat expansion. The heating effect can red shift the emission line of the QD continuously by several $0.1$ nm \cite{gong2010}. During this process, a solid etalon (see Figure 1), which is calibrated by the pump laser of the quantum memory (at $879.7$ nm), is used to set the target wavelength of the QD emission. The FSR and the bandwidth of the etalon are $50$ GHz and $700$ MHz, respectively. The position of the etalon's transmission line can be adjusted by tuning the temperature of the etalon. When the temperature is set to $38.1$ $^{\circ}$C, the etalon's transmission line is well calibrated to $879.7$ nm. The transmission efficiency is greater than $95\%$. The wavelength of the single photons can be finely tuned to the etalon's transmission line by changing the power of the heating laser to $24$ mW.

\textbf{Quantum memory sample.} The sample that we use for the quantum memory is composed of two pieces of nearly identical Nd$^{3+}$:YVO$_{4}$ crystals with a $45^{\circ}$ half-wave plate sandwiched between them (see Figure 1). Each crystal is $3$ mm long along the $a$ axis and has a doping level of $5$ ppm \cite{zhouLGI}. This sample is cooled to $1.5$ K in a liquid-helium-free cryostat (Oxford Instruments, SpectromagPT) with a superconducting magnetic field of $0.3$ T parallel to the crystals' $c$ axis (the axes of these two crystals are parallel). The $^{4}$I$_{9/2}\rightarrow^{4}$F$_{3/2}$ transition (approximately $879.7$ nm) of the crystals is used for the quantum memory here, as mentioned earlier. In this transition, the absorption of the $|H\rangle$-polarized photons in the crystals is much stronger than that of the $|V\rangle$-polarized photons, with ``$H$" (``$V$") denoting the horizontal (vertical) direction, i.e., the direction parallel (perpendicular) to the crystals' $c$ axis. This sandwich-like structure has been demonstrated to have the ability of maintaining the polarization characteristic of the stored photons, thus enabling the high-fidelity memory of the polarization-based qubit \cite{zhou2012}.

\textbf{Time sequence and quantum memory setup.}
The time sequence is shown in Figure 3(a). During the $11.5$-ms preparation time, the frequency of the pump laser ($|H+V\rangle$-polarized so that both crystals can be pumped, see the $879.7$-nm laser in Figure 1) is modulated by AOM2 and an EOM to generate a comb-like profile with a bandwidth of approximately $500$ MHz and a programmable comb period $\Delta$ (see Supplementary Note 4 for details). Next, this pump laser is passed through Chopper1, and then focused on the Nd$^{3+}$:YVO$_{4}$ crystals by Lens2 ($f=250$ mm) to prepare the $500$-MHz-width AFC. Subsequently, the laser is collimated by Lens3 ($f=250$ mm) and blocked. Meanwhile, an anti-phase chopper (Chopper2) is used to block the residual pump laser to protect the SSPD (Figure 1). After a $2.5$-ms wait time, the $10$-ms storage and retrieval procedure begins. The pump laser is blocked by Chopper1, and single photons come out. The modulation of single photons is achieved by modulating the $633$-nm excitation laser using AOM1 (Figure 1). This method can reduce the loss of single photons. The storage and retrieval period is denoted as $T_{\rm period}$, and the width of the modulated excitation laser pulse is denoted as $T_{\rm expw}$. The single photons hence become a series of pulses as well, with a period of $T_{\rm period}$ and a pulsewidth slightly larger than $T_{\rm expw}$ due to the QD's lifetime. The single photons are then focused on the Nd$^{3+}$:YVO$_{4}$ crystals by the same Lens2 and in the same site as the pump laser to be absorbed by the frequency-tailored ions. The spot size of the pump laser is tuned to $4$ times of that of the single photons, creating good spatial overlap of these two light beams in the non-collinear configuration. This experimental setting sufficiently reduces the noise while maintaining the memory efficiency. Next, after a time of $T_{\rm storage}$, which is determined by the period of the frequency comb, the retrieved single photons are re-emitted as a result of the collective interference among all of the ions that are in phase. Finally, the single photons are collimated by Lens3 and detected by the SSPD, which is immersed in liquid helium and operated at $1.5$ K. Its detection efficiency for our single photons is approximately $8\%$, and the dark count rate is roughly $1.5$ s$^{-1}$. The time spectra are obtained by sending the SSPD signals to a time-correlated single photon counting (TCSPC) system. The electrical pulses that are synchronized with the excitation light pulses ($633$ nm) are used as the trigger signal. The integration range was chosen to be the same as $T_{\rm period}$, and the time interval was chosen to be $1$ ns for Figure 3(b) and $0.5$ ns for Figure 4.

PBS1, HWP1, PBS2, HWP2 and the phase plate are used for the polarization-qubit-memory experiment, the configurations of which have been introduced in the Results section. HWP2 is rotated to the angle where the photon count is maximized when Figure 3(b) and Figure 4 are measured. HWP1 does not affect the maximum value of the photon count, but affects that at which angle of HWP2 the maximum count will appear.

\textbf{Brief summary about the SNR improvement.} First, the QD is grown in a $4\lambda$ planar DBR microcavity. The vertical mode of this microcavity is centered on approximately $879.7$ nm (see Supplementary Note 1 and Supplementary Note 2 for details). This structure makes the QD preferentially emit vertically, where the single photons are collected, in contrast with the $4\pi$-solid-angle emission corresponding to a free QD \cite{flagg2009}. Second, we use a high N.A. ($0.68$) aspheric lens (Lens1) placed in the cryostat to collect the single photons instead of a long working distance objective, which has a typical N.A. of approximately $0.4$. The photon count is approximately $5\times10^{5}$ per second with continue-wave excitation. Third, the intensity modulation is performed on the excitation laser instead of on the single photons directly. This method prevents signal loss during the generation of the time sequence. Fourth, in the quantum memory experiment, we use a non-collinear configuration for the pump light and the signal photons. This configuration prevents the single-photon loss that occurs during the combination of the signal light beam and the pump light beam, which is required in the collinear configuration. Moreover, the spot size of the focused pump light is $4$ times that of the single photons, allowing the memory efficiency in the non-collinear configuration to be as high as that in the collinear configuration. The non-collinear configuration also greatly suppresses the noise induced by the pump laser. Finally, an SSPD is used instead of a silicon single-photon avalanche diode. The dark count rate of the SSPD can be as low as $1.5$ s$^{-1}$.

{\bf Acknowledgements}\\
This work is supported by the National Natural Science Foundation of China (Grants Nos. 61490711, 11274289, 11304305, 11325419, 61327901 and 91321313), the National Basic Research Program of China (Grants Nos. 2011CB921200 and 2013CB933304), the Strategic Priority Research Program(B) of the Chinese Academy of Sciences (Grant No. XDB01030300), the Fundamental Research Funds for the Central Universities (No. WK2470000011), the China Postdoctoral Science Foundation funded project (Grant No. 2012M521229).

{\bf Author Contributions}\\
C.-F.L., J.-S.T. and Z.-Q.Z. plan and design the experiment. J.-S.T. carries out the quantum-dot part of the experiment assisted by Y.-T.W. and G.C.. Z.-Q.Z. carries out the quantum-memory part of the experiment assisted by X.L. and Y.-L.H.. Y.Z. and Y.-N.S. design the computer programs. S.W. and D.-Y.H. operate the SSPD. Y.Y., M.-F.L., G.-W.Z., H.-Q.N. and Z.-C.N. grow the quantum dot sample. Y.-L.L. fabricates the quantum dot sample. J.-S.T. and Z.-Q.Z. analyze the experimental results. J.-S.T. writes the manuscript with the help of C.-F.L., Z.-Q.Z. and Y.Y.. G.-C.G. and C-F.L. supervise the project. All authors discuss the experimental procedures and results.

{\bf Competing financial interests}\\
The authors declare no competing financial interests.

\section{Supplementary Figures}
\begin{figure*}[h]
\centering
\includegraphics[width=0.45\textwidth]{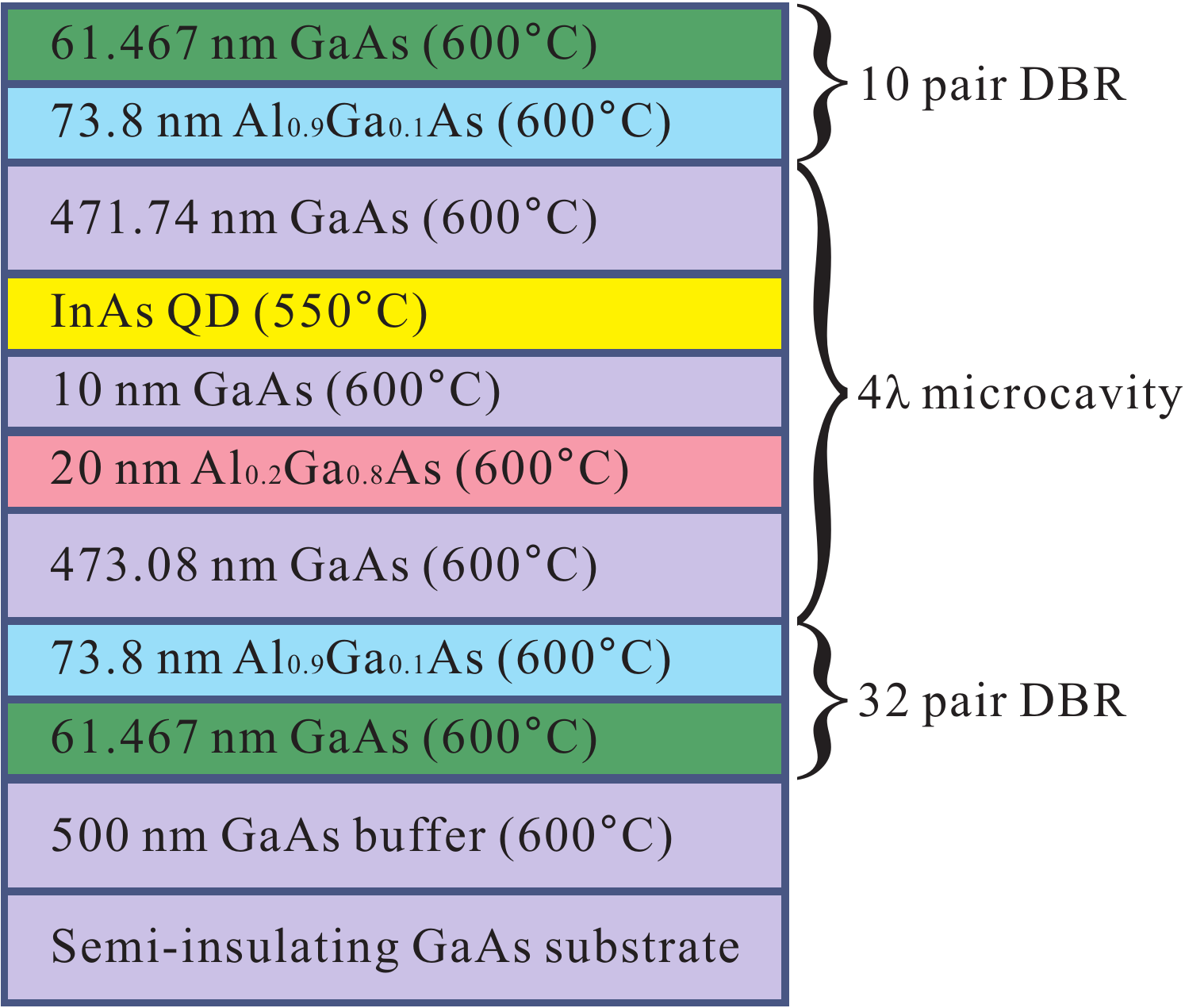}
\renewcommand{\figurename}{Supplementary Figure}
\caption{\label{FigS1} \textbf{The structure of QD sample.} The $20$-nm-thick Al$_{0.2}$Ga$_{0.8}$As layer blue shifts the emission of the QD to approximately $879.7$ nm. The InAs QD is grown in a $4\lambda$ planar DBR microcavity.}
\end{figure*}

\begin{figure*}[h]
\centering
\includegraphics[width=0.9\textwidth]{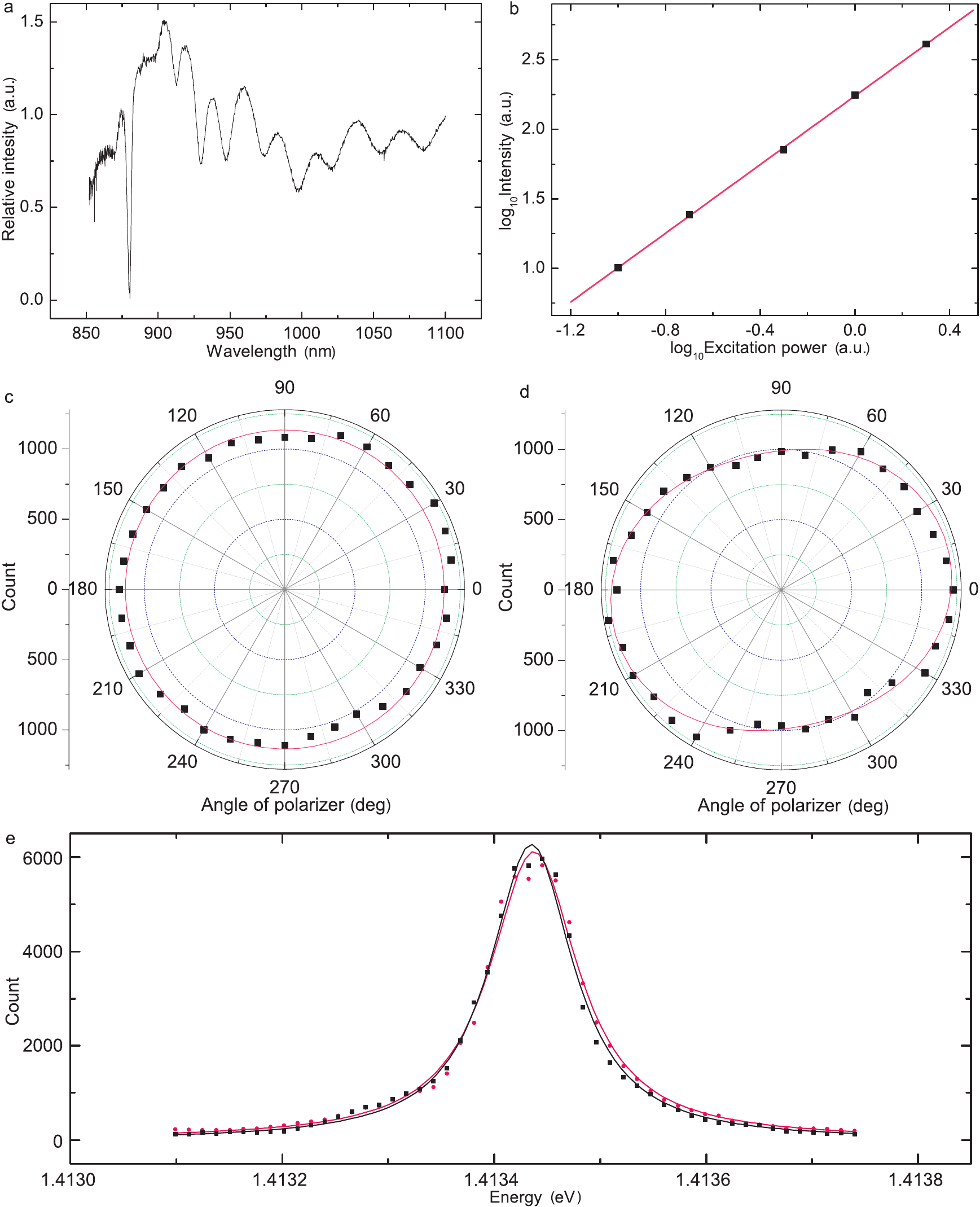}
\renewcommand{\figurename}{Supplementary Figure}
\caption{\label{FigS2} \textbf{Studies of the QD's fluorescent characteristics.} (a) The reflective spectrum of DBR microcavity. The narrow dip at $880$ nm is cavity mode of the DBR microcavity. The FWHM of this dip is $3$ nm, which corresponds to the cavity Q value of $293$. (b) Logarithmic plot of the intensity versus excitation power for the peak in Figure 2(a) in main text. The slope is fitted to be $1.235\pm0.011$. (c, d) The polarization dependent measurement of the QD emissions without (c) and with (d) a QWP inserted before the polarizer. The pink lines are the theoretical fittings, respectively. (e) The polarization-resolved PL spectra of the QD. The black dots are obtained when the polarizer is set to the direction of $\pi_{x}$, and the black line is the Lorentz fit of this spectrum. Similarly, the pink dots are obtained when the polarizer is set to the direction of $\pi_{y}$, and the pink line is the corresponding fit. These two spectra coincide to each other well.}
\end{figure*}

\begin{figure*}[h]
\centering
\includegraphics[width=0.15\textwidth]{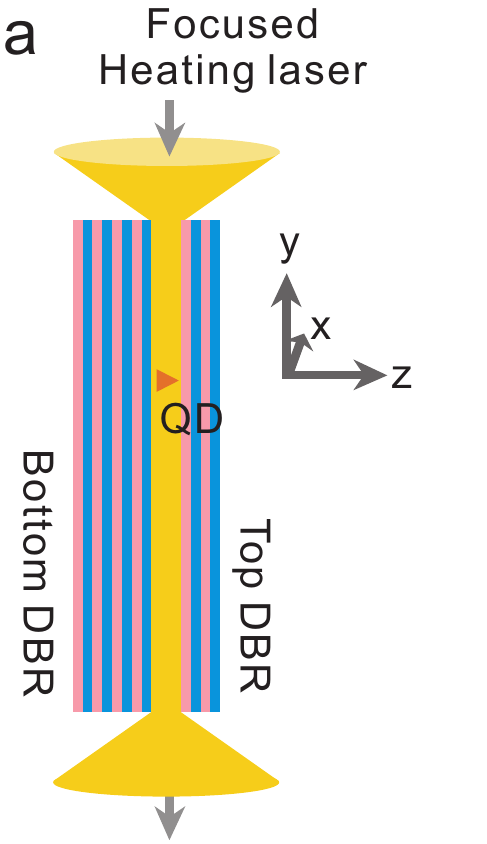}
\includegraphics[width=0.35\textwidth]{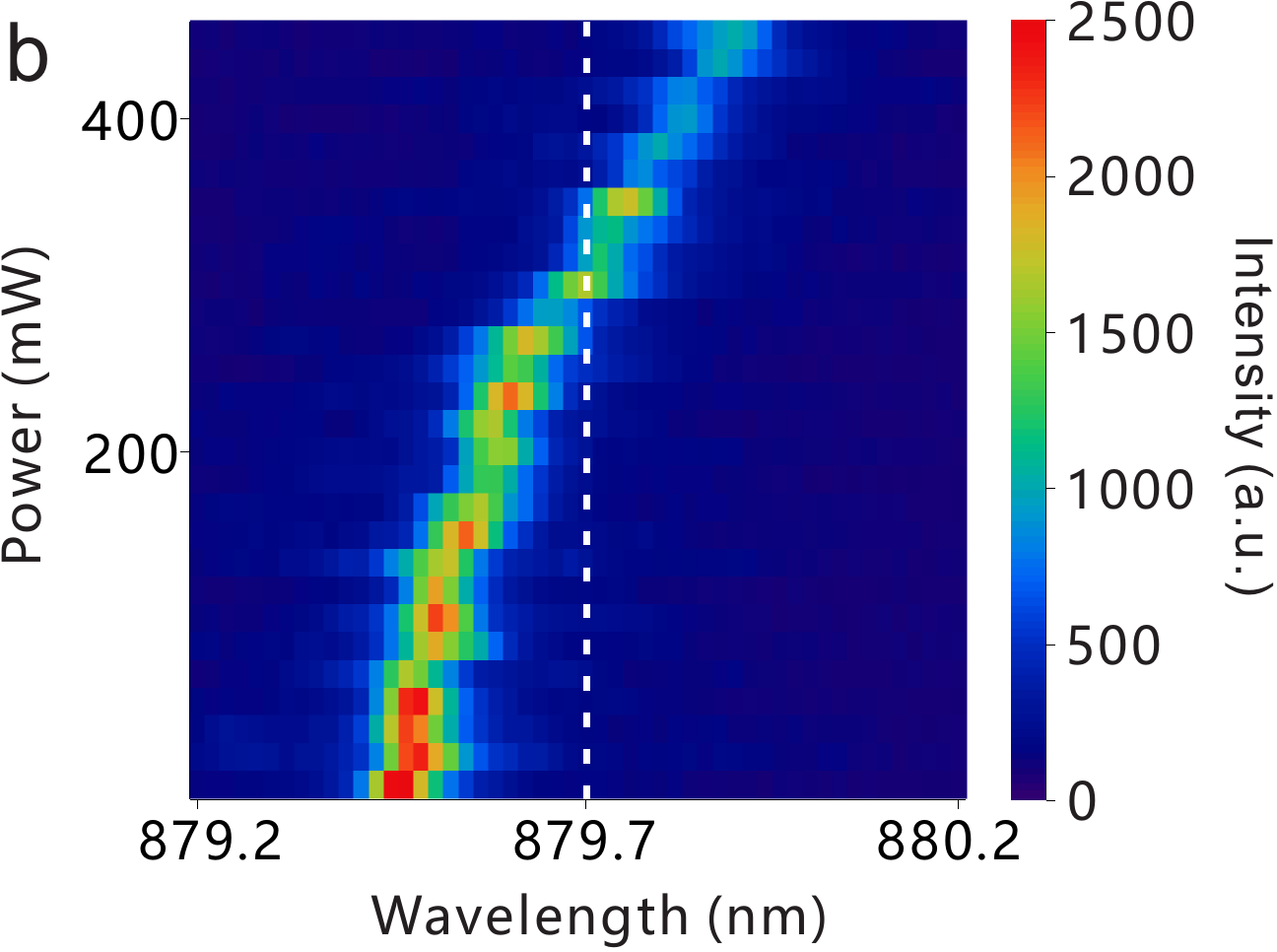}
\renewcommand{\figurename}{Supplementary Figure}
\caption{\label{FigS3} \textbf{The local heating.} (a) The localization of QD area. This is the side view of the QD sample. The heating laser is focused on the cleaved edge of the sample via a long working distance objective. The DBR mirrors here play the role of a waveguide for the heating laser and direct this laser light to the QD to heat the vicinity of the QD. (b) Power-dependent spectra using the $1550$-nm laser. The power of the $1550$-nm heating laser should be much larger than that of the $910$-nm laser. $300$ mW is required to shift the peak to $879.7$ nm.}
\end{figure*}

\begin{figure*}[h]
\centering
\includegraphics[width=0.4\textwidth]{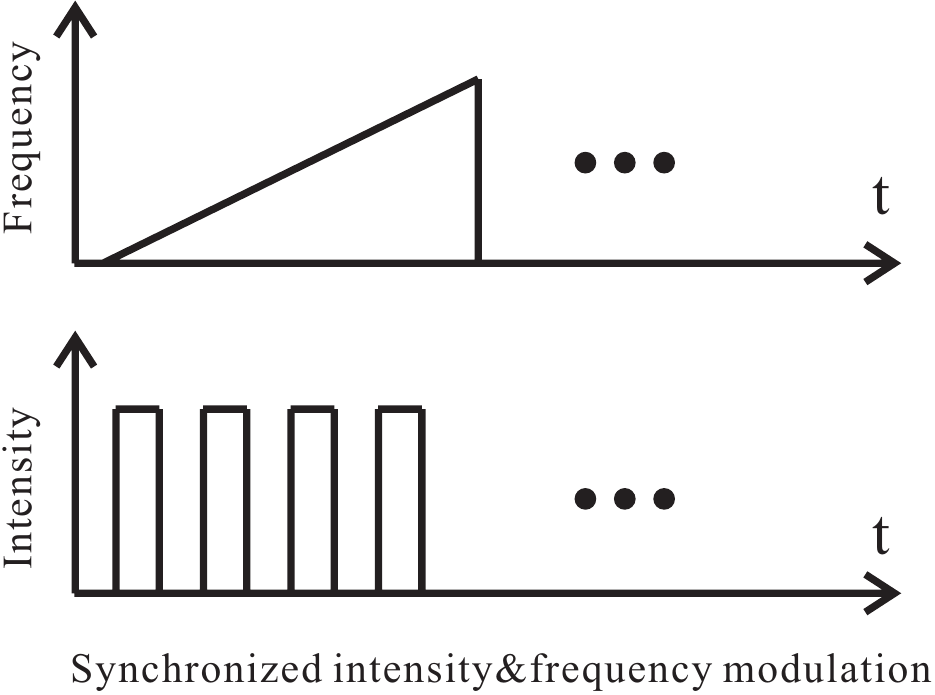}
\renewcommand{\figurename}{Supplementary Figure}
\caption{\label{FigS4} \textbf{The modulation of pump laser.} The intensity of the pump laser is modulated by an AOM, which simultaneously shifts the laser frequency continuously.}
\end{figure*}

\begin{figure*}[h]
\centering
\includegraphics[width=0.4\textwidth]{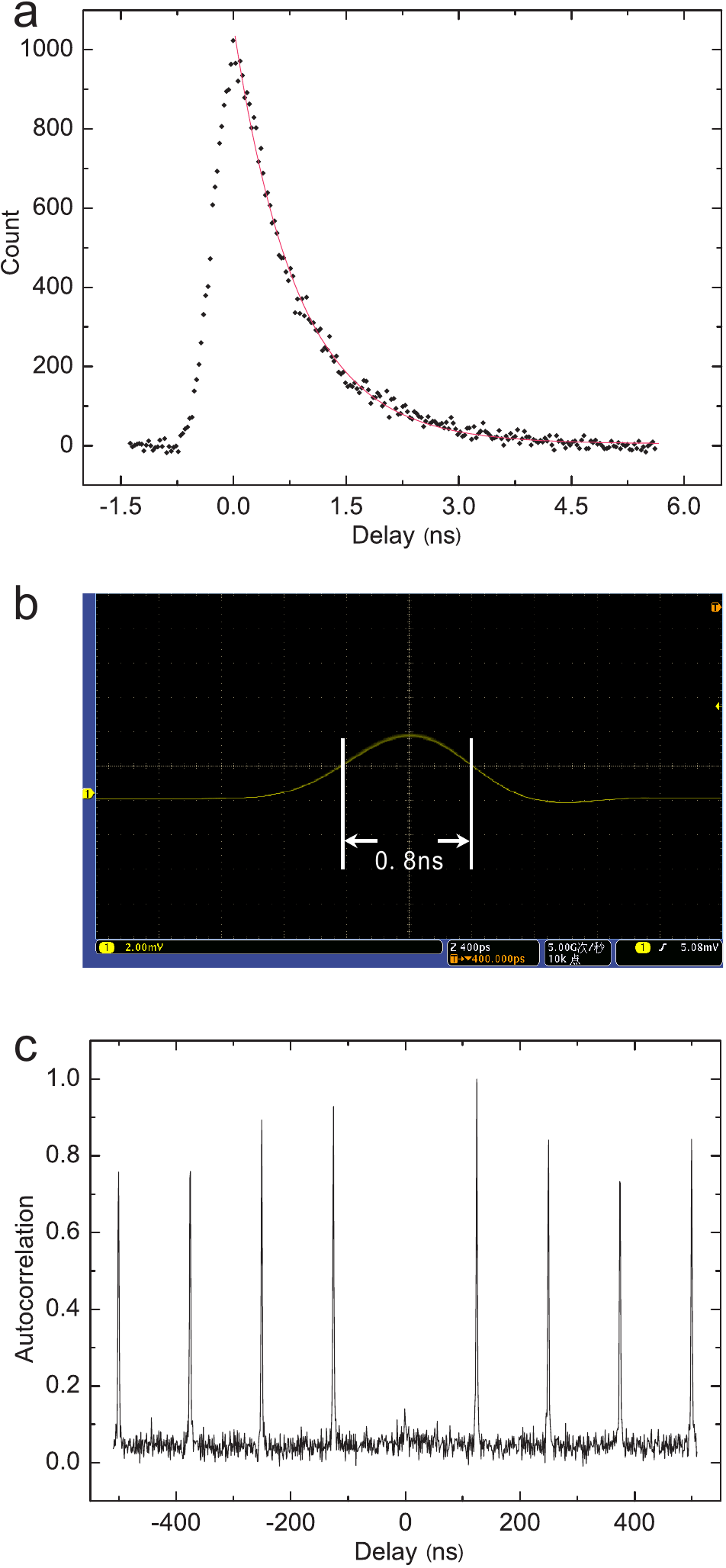}
\renewcommand{\figurename}{Supplementary Figure}
\caption{\label{FigS5} \textbf{The single-photon characteristics.} (a) The lifetime measurement. The QD is excited using a $76$-MHz picosecond pulse laser, and the data are collected and analyzed by a PTA. We use the exponential decay function to fit these data. The pink line is the fitting result, and the QD's lifetime is $0.849\pm0.011$ ns. (b) The temporal width of the excitation pulses. The yellow line is a measurement of the excitation laser chopped by an EOM driven by a series of fast electric pulses, using a fast photodetector and a  fast oscilloscope. The pulsewidth is $0.8$ ns. (c) The HBT experimental result. The missing of the zero-delay peak confirms there is only one photon in each pulse.}
\end{figure*}

\clearpage
\section{Supplementary Notes}

\subsection{Supplementary Note 1}
\textbf{The quantum dot sample.} The quantum dot (QD) sample is prepared on a semi-insulating (100) GaAs substrate with a VEECO Gen-II solid source molecular beam epitaxy (MBE) system. As illustrated in Supplementary Figure 1, the structure consists of a distributed Bragg reflector (DBR) with $32$ periods of $\lambda/4$ Al$_{0.9}$Ga$_{0.1}$As/GaAs as the bottom mirror, a $4\lambda$-thick cavity in the center, and $10$-periods Al$_{0.9}$Ga$_{0.1}$As/GaAs DBR as the top mirror. An InAs QD layer with an ultra-low density of approximately $10^{8}$ cm$^{-2}$ is at the center of the $4\lambda$ planar cavity. This QD layer is grown $10$ nm above the Al$_{0.2}$Ga$_{0.8}$As layer to blue shift the emission wavelength to approximately $879.7$ nm. Note that there are other ways to tune the emission of QDs to $879.7$ nm, for example, using AlGaInAs QDs \cite{schlereth2008} or nonlinear frequency conversion ($35\%$ to $40\%$ external conversion efficiency) \cite{ates2012}. We just have used the method that is more suitable for our experiment.

\subsection{Supplementary Note 2}
\textbf{The photoluminescence spectrum.} As shown in Figure 2(a) in the main text, there is a single peak in the photoluminescence spectrum. Two reasons may explain this result. One reason may be that the other peaks are filtered or weakened by the DBR microcavity. The reflection spectrum of the DBR microcavity is shown in Supplementary Figure 2(a). To derive this spectrum, a thermal light source with a wide spectrum is used to irradiate a mirror and the sample, respectively, and two spectra are obtained. Next, the latter spectrum is divided by the former spectrum to obtain the relative-intensity spectrum. In this figure, the background is subtracted. We find that a sharp dip appears at the position of $880$ nm. This dip shows the cavity mode. When the light is resonant with the cavity mode, it is more likely to transmit through the DBR cavity, resulting in the dip in the reflected signal. The full width of half maximum (FWHM) is $3$ nm, so other peaks are very likely to be weakened, whereas this peak will be definitely strengthened by the DBR cavity because of the improvement of the collection efficiency.

Supplementary Figure 2(b) shows the logarithmic plot of the intensity versus excitation power for this single peak. The pink line is the linear fit for the data. The slope is fitted to be $1.235\pm0.011$. According to the rate equations of the carriers in QD, the populations of the exciton and biexciton will increase linearly and quadratically, respectively, as the excitation power increases. Therefore, the slope should be $1$ for the exciton and $2$ for the biexciton. The intermediate value indicates that this peak is probable to be a trion \cite{besombes2002,rodt2005,gomisbresco2011}. This may be the second reason because the exciton line and biexciton line always appear as a pair, but the trion could have only a single line. The reason for this behavior is that there are two pairs of hole-electrons in a biexciton but only one pair in a trion (and an additional electron or hole). In the experiment, the generation of the trion is possible. Because the mass of the hole is much larger than that of electron, when they are excited, the electron moves faster than the hole. This unbalanced carrier capture enables a trion to be produced more easily.

We also perform the polarization dependent emission measurements for this QD using a quarter-wave plate (QWP) and a polarizer. The polarization state of the trion emission is supposed to be a mixture of $|\sigma^{\pm}\rangle=\frac{1}{\sqrt{2}}(|H\rangle\pm i|V\rangle)$, where $|H\rangle$, $|V\rangle$ denote the horizontal and vertical polarization states, respectively. Therefore, when only a polarizer is used in the measurement, the polarization spectrum is calculated to be a constant $A$; and when a QWP is inserted before the polarizer, the polarization spectrum is derived as $A_{1}cos^{2}(\theta-\theta_{0})+A_{2}sin^{2}(\theta-\theta_{0})$, with $\theta$ representing the polarizer's angle and $\theta_{0}$ associated to the QWP's angle. The experimental results of these two situations are shown in Supplementary Figure 2(c) and Supplementary Figure 2(d), respectively, which are then well fitted with the corresponding theoretical formulas. The fitting results show the ratio $A_{2}/A_{1}$ to be $0.801\pm0.015$.

Furthermore, we measure the polarization-resolved photoluminescence (PL) spectra of this peak using a linear polarizer and a piezo driven etalon with a free spectral range (FSR) of $300$ GHz and a bandwidth of $8$ GHz. The results are shown in Supplementary Figure 2(e). The black dots and the pink dots correspond to single-photon emissions detected when the polarizer is set to the directions of $\pi_{x}$ and $\pi_{y}$ (the definitions of these notations can be referred in \cite{bayer1999}), respectively. If the QD contains an exciton or a biexciton, the exchange interaction will cause a fine structure splitting (FSS) between these two polarization-resolved spectra; whereas, if the QD contains a trion, the exchange interaction will vanish, so the FSS will also vanish \cite{bayer1999,bayer2002}. Here by fitting our results respectively with Lorentz function \cite{tang2012epl} (the black and pink lines in the figure corresponding to the data represented with the same-color dots, respectively), we derive the difference of the centers of these two polarization-resolved spectra, which is $2.277\pm1.383$ $\mu$eV. This value is much less than the FSS reported in the previous works (typically tens of $\mu$eV to greater than $100$ $\mu$eV) \cite{bayer1999,bayer2002,ramirez2010}. Therefore, to some extent we can conclude that our QD contains a trion within the accuracy of the experiment. The non-vanished value of the splitting may be caused by the imperfect optical mirrors and windows of the cryostat.

Although we suppose that our QD may contain a trion, we note that this is not an essential issue in this work. Exciton, biexciton and trion are equivalent here since they all emit true single photons. With this property at hand, reliable quantum repeaters can be constructed \cite{sangouard2007}. Furthermore, when specific carriers are used in our configuration, more advanced quantum applications will be realized.

\subsection{Supplementary Note 3}
\textbf{The local heating.} Supplementary Figure 3(a) shows a sketch of how the heating laser is directed to the vicinity of the selected QD (side view of the sample). First, the laser is focused to the cleaved edge of the QD sample using a 50X long working distance ($20$ mm) objective with N.A. of $0.42$. Next, the objective is moved along the $z$ axis until the DBR microcavity is found. In this situation, the DBR mirrors acts as a waveguide for the heating laser with the $z$ direction restricted \cite{flagg2009}. The separation between these two DBR is approximately $1$ $\mu$m. Meanwhile, the dispersion in the $x$ direction is estimated to be less than $0.4$ mm due to the high refractive index (approximately $3.3$) of the sample and considering the height (along $y$ axis) of the sample of $3$ mm. Next, we move the objective along $x$ axis to find the selected QD with the wavelength-shift effect becoming closer to optimal. Finally, the distance between the objective and the sample in $y$ direction can also be tuned slightly to make the focused spot be closer to the QD. Notably, other QDs exposed by this laser are also heated, but because the volume of the heating area is very small, the global sample remains cool; therefore, the inconvenience caused by heat expansion can still be avoided.

In this experiment, a high-power $910$-nm laser was used as the heating laser, which has a photon energy below the bandgap of GaAs. However, the sample does not appear absolutely transparent for this light. There exists a weak absorption which may be due to the imperfections of the material of the sample, for example, the defect centers generated during the sample growth, which can absorb the heating photons and convert the energy into phonons. This phenomenon also appears in other systems, for example, the absorption of $1000\sim1600$-nm light in a silicon spherical microcavity \cite{garin2014}. In our situation, the laser power is high ($24$ mW); therefore, we can observe the heating effect clearly. We also use a $1550$-nm laser to repeat the local-heating process, with the result shown in Supplementary Figure3(b). We find the power of the laser should be much larger than that in the $910$-nm case. Approximately $300$ mW is required to shift the peak to $879.7$ nm, whereas the ratio of the coupling efficiencies to the DBR waveguide for the $910$-nm and $1550$-nm lasers is estimated to be not more than $2:1$. This result can be explained by the fewer defect centers that are able to be excited by the $1550$ nm photons because of the smaller photon energy.

There are other technologies to finely tune the wavelength of the QD's emission, such as electric field, strain, or magnetic field. Compared with these methods, local heating is more suitable for us. The first reason is that local heating is simple. Only a high-power laser is required, without the complex sample fabrication procedures, the requirement of piezoelectric actuators placed in the cryostat, or the use of a superconducting magnet. The second reason is the DBR mirrors that are used for improving the collection efficiency of single photons are just right used as a waveguide for the heating laser to flow to the selected QD.

\subsection{Supplementary Note 4}
\textbf{The preparation of the atomic frequency comb.} The atomic frequency comb (AFC) technique is utilized to realize reversible quantum state transfer between single photons and atomic excitations \cite{AFC08,AFC09}. The AFC protocol requires a tailored absorption profile composed of a series of periodic and narrow absorbing peaks separated by $\Delta$. The input single photons can be absorbed and diffracted by the AFC. A collective echo emission is retrieved after a storage time $T_{\rm storage}=1/\Delta$ in the crystals.

The pump light for establishing AFC comes from a frequency stabilized Ti:sapphire laser (MBR-110, Coherent). Using the double-pass configured acousto-optic modulator (AOM2 in Figure 1 of main text), the pump laser frequency is swept over $100$ MHz in each $200$ $\mu$s cycle, and its amplitude is modulated periodically to give a comb structure (Supplementary Figure 4). This pump sequence is continuously repeated in the preparation phase to produce an optimized AFC structure. The AFC bandwidth prepared with the AOM is approximately $100$ MHz and is limited by the effective bandwidth of the double-pass configured AOM. To further extend the memory bandwidth, the output light of the AOM is directed into a fiber-coupled high-speed electro-optic phase modulator (EOM in Figure 1 of the main text, where the fibers are not sketched) that creates first and second-order sidebands \cite{eomband}. The comb at the carrier frequency is copied twice on each side. To ensure zero frequency detuning for the copied combs, the driver frequency for the EOM is chosen as $100$ MHz. The radio frequency (RF) power for driving the EOM is experimentally optimized to optimize the storage efficiency. In this way, an overall comb width of approximately $500$ MHz is achieved in our experiment. During the $11.5$-ms preparation time, the pump laser is sent to the Nd$^{3+}$:YVO$_{4}$ crystals to prepare the AFC by pumping ions with certain frequencies to the auxiliary state (see the inset of Figure 1 in the main text). The lifetime of the ground state and the auxiliary state is $42.9$ ms, much longer than the $10$-ms storage-and-retrieval time. This feature ensured that the single photons can be successfully stored in the storage-and-retrieval procedure when the pump laser is turned off in order to reduce the pump noise.

\subsection{Supplementary Note 5}
\textbf{The spectral, temporal and photon correlation characteristics of the input and output photons.} Here we primarily discuss the situation of narrow single-photon pulses, which attracts more interest in this work. The linewidth of the QD emission is approximately $25$ GHz, which is determined using a piezo tunable etalon with a FSR of $300$ GHz and a bandwidth of $8$ GHz. The lifetime of this emission is $0.849\pm0.011$ ns, which means its natural linewidth is less than the detected linewidth. This lifetime is derived by using a picosecond pulse laser to excite the QD, and a picosecond time analyzer (PTA) to obtain the time spectrum, which is then fitted with the exponential decay function (see Supplementary Figure 5(a)). The line broadening was mainly caused by spectral diffusion. Resonant excitation can be used to reduce the linewidth in the future works (At best, $7$-MHz linewidth of the QD emission has been achieved using resonant excitation \cite{matthiesen2012}, which is much narrower than the memory bandwidth here). After passing through the filters and the etalons (see Figure 1 in main text), the linewidth of the single photons is reduced to $700$ MHz, which is determined by the bandwidth of the etalons. The memory bandwidth is $500$ MHz; therefore, the linewidth of the retrieved single photons is limited to $500$ MHz. The memory bandwidth is possible to be extended to several $GHz$ by using one AOM and more EOMs to prepare the AFC. The temporal widths of the input and output photons can be derived by fitting the first peak (transmitted photons) and the second peak (stored photons) in Figure 4(a) in main text, respectively, with Gaussian function. The fitting results are $1.478\pm0.128$ ns and $2.400\pm0.081$ ns for the first and second peaks, respectively. The temporal width of the stored single-photon pulse is approximately $1.6$ times of that of the input pulse. This ratio is approximately equal to that of the linewidths of the input photons over the output photons ($700$ MHz$/500$ MHz$=1.4$).

Supplementary Figure 5(b) shows the screenshot of a fast oscilloscope (Tektronix, DPO4104B) with a resolution of $80$ ps. The yellow line is the signal of the excitation laser detected by a fast photodetector (Thorlabs, DET02AFC) with $1.2$ GHz bandwidth and $50$ ps rise time. The excitation laser is modulated to narrow pulses using a $10$ GHz EOM driven by a synchronized high-speed pulse generator (Tektronix, AWG7082C), which supplies a series of $0.5$-ns electric pulses. The FWHM of the excitation pulse is measured to be $0.8$ ns (the difference may be caused by the imperfect cables), which is less than the QD's lifetime of $0.849\pm0.011$ ns. Therefore, the QD can be excited only once per pulse, i.e., only one photon exists in the pulse. The Hanbury Brown-Twiss (HBT) experiment also illustrates this point, as shown in Supplementary Figure 5(c). During this HBT experiment, the electric pulses are supplied with the temporal width of $0.5$ ns (as it is in the memory experiment) and the period of $125$ ns. The integration time is $16.6$ hours. The missing peak at the zero delay demonstrates there is only one photon per pulse. The autocorrelation is estimated to be $0.14$. In the output pulses, there will also be one photon per pulse at most because the memory process will not increase the photon number.

\end{document}